\documentclass[twocolumn,showpacs,amsmath]{revtex4}
\usepackage{graphicx}%
\usepackage{dcolumn}
\usepackage{amsmath}
\usepackage{here}

\begin{document}


\title{Square vortex solitons with a large angular momentum}

\author{Humberto Michinel, Jos\'e R. Salgueiro and Mar\'{\i}a J. Paz-Alonso}

\affiliation{\'Area de \'Optica, Facultade de Ciencias de Ourense,\\ 
Universidade de Vigo, As Lagoas s/n, Ourense, ES-32005 Spain.}

\begin{abstract}
We show the existence of square shaped optical vortices with a large 
value of the angular momentum hosted in finite size laser beams 
which propagate in nonlinear media with a cubic-quintic nonlinearity. 
The light profiles take the form of rings with sharp boundaries 
and variable sizes depending on the power carried. Our stability 
analysis shows that these light distributions remain stable when propagate, 
probably for unlimited values of the angular momentum, provided the hosting 
beam is wide enough. This happens if the peak amplitude approaches a 
critical value which only depends on the nonlinear refractive index of 
the material. A variational approach allows us to calculate the main 
parameters involved. Our results add extra support to the concept of surface 
tension of light beams that can be considered as a trace of the existence 
of a liquid of light.
\end{abstract}

\pacs{46.65.Jx, 42.65.Tg}

\maketitle

\section{Introduction}
In wave mechanics, a vortex is a screw phase dislocation, or defect 
\cite{nye74} where the amplitude of the field vanishes. The phase around 
the singularity has an integer number of windings, $l$, which plays the 
role of an angular momentum. For fields with non-vanishing boundary conditions,
this number is a conserved quantity and governs the interactions between 
vortices as if they were endowed with electrostatic charges\cite{rozas97}. 
Thus, $l$ is usually called the ``topological charge'' of the defect. 

Vortices are present in very different branches of physics like fluid
mechanics, superconductivity, Bose-Einstein condensation, astrophysics 
or laser dynamics, among others\cite{pismen99}. In Optics, a vortex 
with charge $l$ takes the form of a black spot surrounded by a 
light distribution. Around the dark hole, the phase varies from zero 
to $ 2\pi l$. These defects appear spontaneously in light
propagation through turbulent media and can also be produced by appropriately 
shining a computer generated hologram\cite{heckenberg92}. The trace of vortices
in a light field is a characteristic ``fork-pattern'' interferogram 
produced by superposition with a tilted planar wave.

The first experimental works on optical wavefront dislocations were carried 
out in the 80's, in the context of adaptive systems, where phase singularities 
were a severe problem for image reconstruction techniques\cite{baranova8}. 
Later on they have been studied, among other fields, in optical 
tweezing\cite{simpson97}, particle trapping\cite{gahagan96}, 
laser cavities\cite{weiss99}, optical interconnectors\cite{scheuer99} or 
even to perform N-bit quantum computers\cite{mair01}. 

Concerning light vortices in the nonlinear regime\cite{kivshar01},
the first theoretical work analysed their stability in Gaussian-like
distributions propagating in optical Kerr materials\cite{kruglov85}.
It was found for a cubic self-focusing refractive index, that a beam
of finite size will always filament under the action of a phase dislocation. 
This also applies to saturable self-focusing nonlinearities\cite{firth97}.
On the other hand, vortex states were predicted and found experimentally
for self-defocussing materials both in the Kerr case for continuous
background\cite{swartzlander92} and in the saturable case with finite
size beams\cite{tikhonenko96}.

It was shown in \cite{quiroga97} that stable vortex states with $l=1$
can be obtained as stationary states of the propagation of a laser beam through
cubic-quintic optical materials\cite{piekara97,josserand97,davydova04}. 
This kind of nonlinearity is characterized by the $\chi^{(3)}>0$ and 
$\chi^{(5)}<0$ components of the nonlinear optical susceptibility and changes 
from self-focusing to self-defocusing at a given intensity\cite{fang02}. 
It has been recently shown that a gas-liquid phase transition takes place in 
light beams propagating in this type of materials\cite{michinel02}. 

In this work, we will show that stable vortex states with a huge 
value of the angular momentum exist and their peak amplitude and propagation 
constant tend asymptotically, as the beam flux is increased, to 
values that do not depend on $l$. In this way, our results are in 
contradiction with previous work\cite{towers01}, where it was claimed 
that stable vortex states in finite size beams exist only for the 
values $l=1,2$. For $l=3$ it was found a persistent weak instability 
which was also supposed to exist for higher values of the angular momentum.  

In next section we will analyze the cubic-quintic nonlinear model, finding 
numerically the stationary states for a wide range of the angular momentum 
$l$ (up to 50) and describing their particular properties. 
Then, we will calculate analytically, by means of the variational
method, the critical values of the propagation constant and peak amplitude
that characterize the domain of existence of vortices. Finally, we
will perform an azimuthal stability analysis to determine the domain 
zone where stable states can be found.

\section{The model} 
Let us start by writing the equation for laser beam propagation along $z$ 
in an optical cubic-quintic material. 
For paraxial propagation, the equation for the beam envelope $\Psi$ is a 
nonlinear Schr\"odinger equation (NLSE) of the form:

\begin{equation}
\label{NLSE}
i\frac{\partial \Psi }{\partial z}+\nabla _{\perp }^{2}\Psi 
+\left( n_{2}|\Psi |^{2}-n_{4}|\Psi |^{4}\right) \Psi =0,
\end{equation}
 where $\nabla ^{2}_{\perp }=r^{-2}\partial ^{2}/\partial \phi ^{2} + r^{-1}\partial /\partial r+\partial ^{2}/\partial r^{2}$
is the transverse Laplacian operator in cylindrical coordinates 
$(r,\phi,z)$.
The real positive constants $ n_{2}$ and $ n_{4}$ are given 
respectively by the $ \chi ^{(3)} > 0 $ and $ \chi ^{(5)} < 0 $  
components of the nonlinear optical susceptibility and characterize
the dependence of the refractive index on the intensity of the beam.
If $ n_4\leq 0 $, a Gaussian beam of high enough power will undergo 
collapse after self-focusing\cite{chiao64}. 
The effect of a negative fifth order susceptibility ($ -n_{4}$ term) 
combined with diffraction will stop the collapsing tendency for high 
powers, yielding to a stable two-dimensional condensed state of light 
with surface tension properties similar to those of usual 
liquids\cite{piekara97, josserand97, michinel02}.

\begin{figure}
\centerline{\includegraphics[width=3.2 in]{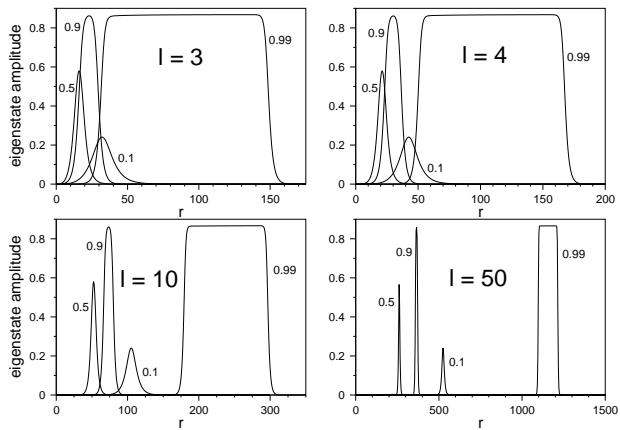}}
\caption{\label{fig1}
Numerically calculated radial amplitude 
profiles of the stationary states of Eq. (\ref{stationary_eq})  
for $l=3$, 4, 10 and 50 
with $ \beta /\beta _{cr}$=0.1, 0.5, 0.9 and 0.99. In all the cases 
$n_{2}=n_{4}=1$ .}
\end{figure}

We are interested in stationary states with radial symmetry and angular 
momentum $l$ of the form:

\begin{equation}
\label{stationary_state} 
\Psi (r, \phi, z)=\psi(r)e^{i l \phi}e^{i\beta z},
\end{equation}
where $\beta$ is the nonlinear phase shift or propagation constant 
and $\psi(r)$ is the radial envelope of the field. After substitution 
of (\ref{stationary_state}) in (\ref{NLSE}), the following $z$-independent 
equation is obtained for $\psi(r)$:

\begin{equation}
\label{stationary_eq} 
-\beta \psi+\nabla_{r}^{2}\psi-\frac{l^2}{r^2}\psi 
+n_{2}\psi^3-n_4\psi^5=0,
\end{equation}
where $\nabla^2_{r}\equiv \partial^2/\partial r^2 + 
(1/r)\partial /\partial r$, is the radial part of the Laplace operator. 

For a given integer value of $l$, a continuum of eigenstates 
with $\psi\rightarrow 0$  as $r\rightarrow \infty$ can be 
obtained by solving numerically Eq.(\ref{stationary_eq}). 
Close to the origin the shapes follow the linear regime with 
$\psi \propto r^{l}$.
To this aim, we have used a standard relaxation technique. 
The profiles of the eigenstates for several values of $l$ and
$\beta$ are plotted in Fig. \ref{fig1} for the case of $n_2=n_4=1$. 
 We particularly show states with $l=3$ and $l=4$ since 
these were previously found unstable in previous work\cite{towers01}, 
as well as two examples of large angular momentum states ($l=10$ and $l=50$).
In all cases, the stationary states can only be found for values 
of $\beta$ between zero and a fixed critical value 
$ \beta _{cr}$\cite{quiroga97}, which does not depend on $l$.

It can be appreciated in the graphs that values of $ \beta $ below 
$ 0.5\beta_{cr}$ yield to light distributions with smooth and 
wide Gaussian-like shapes. As $\beta$ is incremented the beam flux 
grows and the spatial profiles narrow, yielding to a minimum thickness 
of the ring of the stationary state for values of $\beta$ around 
$ 0.8\beta_{cr}$, keeping approximately the Gaussian shape. For larger 
values of the propagation constant, the beam flux grows rapidly with $\beta$ 
and the peak amplitude of the distribution saturates due to the effect of 
$n_{4}$, reaching asymptotically the value $ A_{cr}$ which is slightly below 
the maximum amplitude. Thus, high power beams show spatial light 
distributions with flatted tops in their profiles, similar to 
those of hyper-Gaussian functions\cite{dimitrievski99,quiroga99}. 

\begin{figure}[h]
\centerline{\includegraphics[width=3.2 in]{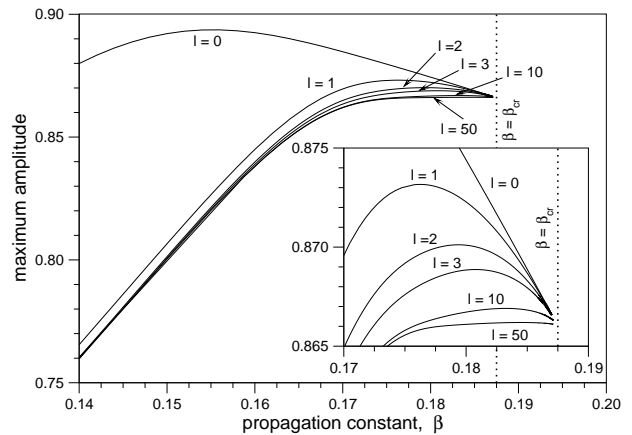}}
\caption{\label{fig2}
Maximum amplitude of the stationary states versus propagation constant 
for angular momenta $l$=0, 1, 2, 3, 10 and 50. All 
the curves join at the point $ A=A_{cr}=0.866$, 
$\beta = \beta _{cr}=0.1875$. Inset: detail of the zone
close to $\beta _{cr}$ where the calculations become delicate. }
\end{figure}

We must stress the intriguing fact that both $ \beta _{cr}$ and $ A_{cr}$
do not depend on the value of the topological charge. This is shown
in Fig. \ref{fig2}, where the maximum amplitude has been plotted
as a function of $ \beta $. In the inset, the zone $\beta\approx\beta_{cr}$ 
can be seen in detail. As can be appreciated, whatever the 
value of $l$ is, all the curves tend to join at the same point. 
This means that the critical value of the propagation constant and peak 
amplitude only depend on the nonlinearity and not on the angular momentum. 
We will revise this result in our analytical study of the next section.

\begin{figure}
\centerline{\includegraphics[width=3.2 in]{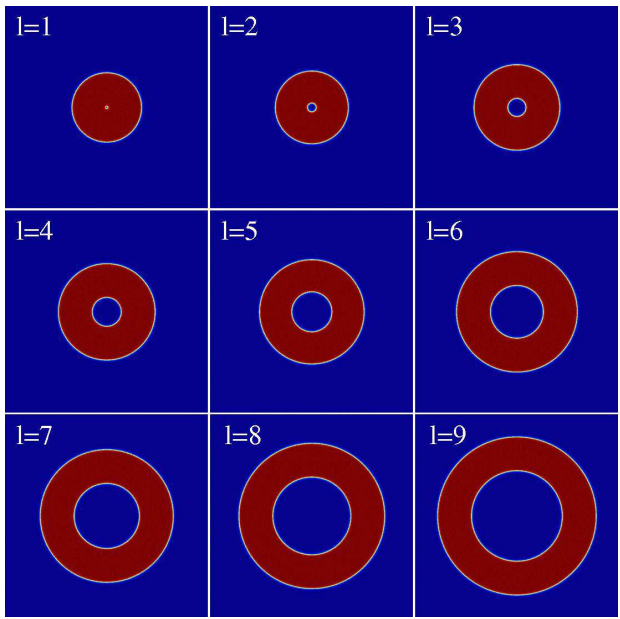}}
\caption{\label{fig3}
(Color on line) Azimuthal eigenstates of Eq.(\ref{stationary_state})
for $l=1$ to $9$ with $\beta=0.95\beta_{cr}$.}
\end{figure}

\begin{figure}
\centerline{\includegraphics[width=3.2 in]{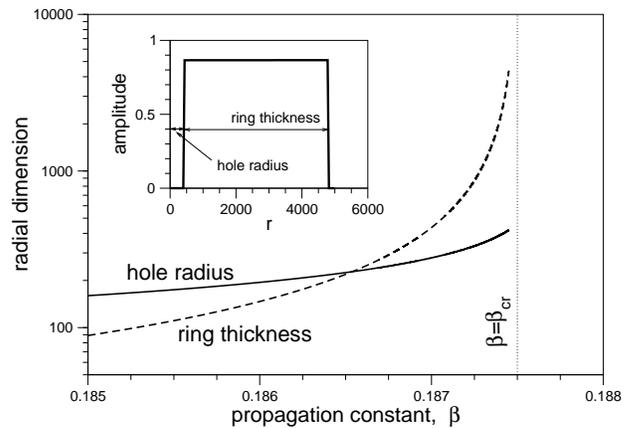}}
\caption{\label{fig4}
Dependence of the internal hole radius (continuous line) and ring 
thickness (dashed line) with the propagation constant $\beta$ in the
vicinity of $\beta_{cr}$ for eigenstates with $l=10$. Inset: 
example of a stationary state with $l=10$ and $\beta=0.9998\beta_{cr}$.}
\end{figure}

It also worths to mention that the central hole increases its size
with the topological charge for a fixed value of $\beta$, as can 
be seen comparing the profiles in Fig. \ref{fig1} for $ l=3,4$ with 
$ l=10,\, 50$. This is also clearly shown in Fig \ref{fig3}, where
we plot several eigenstates with values of the angular momentum ranging 
from $l=1$ up to $l=9$, with propagation constant $\beta=0.95\beta_{cr}$.
Besides, if $\beta$ grows, the radius of the hole increases.
As the value of $\beta$ approaches $\beta_{cr}$ the thickness of the external
ring grows faster than the internal hole, and the final result
takes the asymptotic form of a dark spot surrounded by a larger
ring of light of almost constant shape which ends abruptly at a given
radius. This behavior can be assessed having a look to Fig.\ref{fig4}, 
where the dimensions of the internal hole and the ring thickness are plotted 
versus the propagation constant for the particular case of $l=10$. A 
logarithmic scale was chosen to highlight that the growth in the ring 
thickness clearly dominates over the hole radius from certain value of 
the propagation constant. In the inset it is also shown, as an example, 
one of the stationary states with $\beta$ very close to $\beta_{cr}$, 
showing the huge ring whose width clearly exceeds the hole 
radius and presents a practically rectangular shape. 

\section{Variational analysis}

To explain the properties of the above light distributions, we have
performed a variational analysis\cite{dimitrievski99,quiroga99}. It is easy to 
demonstrate that the stationary system described by the model 
(\ref{stationary_eq}) can be obtained from the following Lagrangian function:

\begin{eqnarray}
\label{lagrangian} 
\mathcal{L}=r\left\{\left(\frac{d\psi}{dr}\right)^2+
\left(\frac{l^2}{r^2}-\beta\right)\psi^2
-\frac{n_2}{2}\psi^4-\frac{n_4}{3}\psi^6\right\}.
\end{eqnarray}
Now assuming a rectangular shape of the stationary 
states for values of $\beta$ close to $\beta_{cr}$
(see Fig.~\ref{fig1}), we choose a trial function 
$\psi_v(r)$ centered at $r_0$ with amplitude $A$ and width $2w$,
given by:

\begin{eqnarray}
\label{trial_function} 
\psi_v(r)=\left\{
	\begin{array}{ll}
	A, & |r-r_0|\le w\\
	0, & |r-r_0|> w
	\end{array}
\right.
\end{eqnarray}
Thus, $r_0$, $A$ and $w$ are the variational parameters. Using this trial 
function and minimizing the average over the Lagrangian $<\mathcal{L}>$ 
respect to each parameter\cite{anderson83}, we obtain the following conditions:

\begin{eqnarray}
\label{variational_1}
\frac{l^2}{r_0^2-w^2}+\frac{n_2}{2}A^2-\frac{n^4}{3}A^4-\beta=0,\\
\label{variational_2}
\frac{l^2}{2r_0w}\ln\left(\frac{r_0+w}{r_0-w}\right)-
n_2A^2-n_4A^4+\beta=0.
\end{eqnarray}
where Eq.~(\ref{variational_1}) is obtained from the minimization respect to 
$r_0$ and $w$ (the same condition is obtained for both parameters), 
and Eq.~(\ref{variational_2}) follows from minimization respect to 
parameter $A$. In the limit when $\beta \rightarrow \beta_{cr}$, the 
first term of both equations vanishes. This follows by taking into account 
that $r_0>w$ should always be satisfied (otherwise there would be no 
hole), then we have $r_0^2-w^2>r_0^2 \rightarrow \infty$ and 
$(r_0^2-w^2)^{-1}\rightarrow 0$, consequently the first term of 
Eq.~(\ref{variational_1}) is zero. For Eq.~(\ref{variational_2}), 
the argument of the logarithm tends to infinity as it is easily deduced 
from the fact that the ring width grows faster than the hole radius and 
$r_0>w$ ($(r_0+w)/(r_0-w)>2w/(r_0-w)\rightarrow \infty$). However, this term 
diverges logarithmically, meanwhile the denominator goes to 
infinity quadratically (product $r_0w$), and consequently the whole term 
tends to zero. Finally, we can solve them for $A_{cr}$ and $\beta_{cr}$ to 
obtain:

\begin{eqnarray}
\label{a_cr}A_{cr}={\left(\frac{3n_2}{4n_4}\right)}^{1/2},\\
\label{beta_cr}\beta_{cr}=\frac{3n_2^2}{16n_4}.
\end{eqnarray}
For the particular case considered in the numerical 
calculations displayed in Figs.~\ref{fig1}-\ref{fig2}, i.e. 
taken $n_2=n_4=1$, the values obtained for the 
critical parameters are  $A_{cr}=0.866025$ and $\beta_{cr}=0.1875$.
The comparison of these analytical results with the numerical
calculations shows an excellent agreement, since both values are 
exactly those guessed numerically (see Fig.~\ref{fig2}). 
This so good result is due to the choice of the trial function, 
which fits almost exactly with the numerical solution for values 
close to the critical point.

\section{Stability analysis} 

In order to test the stability of 
the stationary states we calculated the growth rates of small azimuthal 
perturbations to find out the value of $\beta$ at which they vanish.  
Additionally, in order to assess the accuracy of the previous analysis, we 
propagated some {\em unstable} eigenstates with a split-step Fourier method 
and found their splitting distances. The inverse of these values should 
coincide, except for a constant scale factor, with the dominant perturbation 
eigenvalues calculated in the azimuthal instability analysis. 
Finally, we have also simulated other kind of perturbations like 
total reflection at the boundary between a cubic-quintic material
and air. As we will see below, the eigenstates show robust behavior against
these collisions and preserve their angular momentum although strong
oscillations are observed.

\begin{figure}
\centerline{\includegraphics[width=3.2 in]{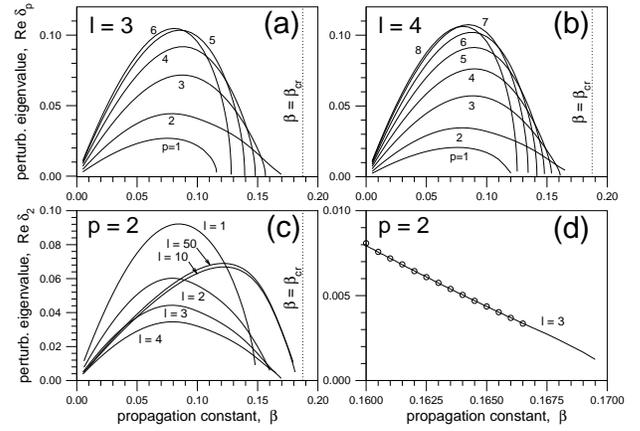}}
\caption{\label{fig5}
Growth rate of the azimuthal perturbation
versus $\beta$. In (a) and (b), the angular momentum 
is fixed ($l=3$ for (a), $l=4$ for (b)) 
and the values of the perturbation order $p$ are 
indicated by labels near the curves. In (c) the growth rate
for the perturbations with $p=2$ for
$ l=1$, 2, 3, 4, 10 and 50 are plotted. In (d), 
a comparison between the inverse of the splitting distance (dots) and 
$p=2$ perturbation eigenvalue for unstable states 
with $l=3$ is shown.}
\end{figure}

To carry out the perturbation analysis, we add to the original eigenstate 
a small $p$-order azimuthal perturbation function\cite{firth97, soto91}:

\begin{equation}\label{perturbed_state}
\tilde \Psi(r,\phi,z) =[\psi(r)+f(r,z)e^{ip\phi}+
h(r,z)e^{-ip\phi}]e^{i(l\phi +\beta z)},
\end{equation}
where $f(r,z)$ and $h(r,z)$ are the small complex components of the 
eigenstate of the $p$-order azimuthal perturbation. Our interest is to 
seek those functions which grow exponentially with $z$, so we assume that
they have the form:

\begin{eqnarray}
\label{perturbation_components}
f(r,z)=[f_1(r)+i f_2(r)]e^{\delta_p z},\\
h(r,z)=[h_1(r)+i h_2(r)]e^{-\delta^{*}_p z},
\end{eqnarray}
being the parameter $\delta_p$ the perturbation eigenvalue. 
In this way, the real part of $\delta_p$ constitutes the growth 
rate of this perturbation. If we replace the perturbed eigenstate 
(Eq.~(\ref{perturbed_state})) into Eq.~(\ref{NLSE}) and keep only 
the first order terms in $f(r,z)$ and $h(r,z)$ (linearisation) 
we obtain the following set of coupled differential equations for 
those components $f(r,z)$ and $h(r,z)$:

\begin{eqnarray}\label{perturbation_system}
i\frac{\partial f}{\partial z}+\nabla^2_{r}f-
\frac{(l+p)^2}{r^2}f+Q(\psi)f+R(\psi)h^*=0\\
i\frac{\partial h}{\partial z}+\nabla^2_{r}h-
\frac{(l-p)^2}{r^2}h+Q(\psi)h+R(\psi)f^*=0,
\end{eqnarray}
where $Q(\psi)\equiv -\beta+(2n_2-3n_4|\psi|^2)|\psi|^2$ and 
$R(\psi)\equiv (n_2-2n_4|\psi|^2)|\psi|^2$.   
The solution of this equation system is obtained using a Crank-Nicholson 
scheme to propagate an initial arbitrary guess until the shape of each 
component does not change perceptibly\cite{soto91}. According to the 
component dependence on $z$ (Eq.~(\ref{perturbation_components})), 
the value of the growth rate can be calculated at each propagation 
step by:

\begin{equation}
\mathrm{Re}\,\delta_p=\frac{1}{2 \Delta z}\ln\frac{|f(r,z+\Delta z)|^2}
{|f(r,z)|^2},
\end{equation} 
where $\Delta z$ is the propagation step and the function $f(r,z)$ is 
evaluated in a fixed point $r$, usually that which corresponds to 
the maximum. Besides, the functions can be rescaled at each step by this 
maximum value to avoid an overflow. The propagation is carried out until 
the value of the perturbation growth rate does not change any more, what 
indicates that convergence was reached. This allows us to obtain the 
growth rates $\mathrm{Re}(\delta_p)$ for different order perturbations 
versus the propagation constant, as depicted in Fig.~\ref{fig5}. 

The growth rates for vortices with angular momentum $l=3$ and $l=4$ are 
shown in Figs.~\ref{fig5}(a)-(b). As can be seen, all of them fall to zero 
for a value of $\beta$ below $\beta_{cr}$, what implies the existence of 
a stability window, in contradiction with previous calculations where all 
the states with $l>2$ where found unstable\cite{towers01}. 
Our results show that the maximum growth rate corresponds to perturbation 
eigenvalues with $p\approx 2l$, what allows to estimate the number of 
filaments $N$ resulting from the breakup of the unstable vortices 
($N\approx 2l$). Besides this, the perturbation $p=2$ has been proved to 
be the most persistent, despite the value of the angular momentum. Hence, 
in Fig.~\ref{fig5}(c) we plot the curves associated to this perturbation 
for different values of the angular momentum, including the cases 
corresponding to $l=10$ and $l=50$. As can be appreciated in these plots,
 there exists a window between the vanishing point and the limit value for 
$\beta$ ($\beta=\beta_{cr}$), what proves the existence of a 
stability zone close to the critical point containing an infinite 
number of stable eigenstates. Note that this window narrows for high 
values of $l$ but remains finite. 
As $l$ increases, the point at which the perturbation vanishes approaches 
asymptotically the critical point. However, we believe that the critical 
point itself is never reached, even for arbitrarily high values of the 
topological charge.

\begin{figure}
\centerline{\includegraphics[width=3.2 in]{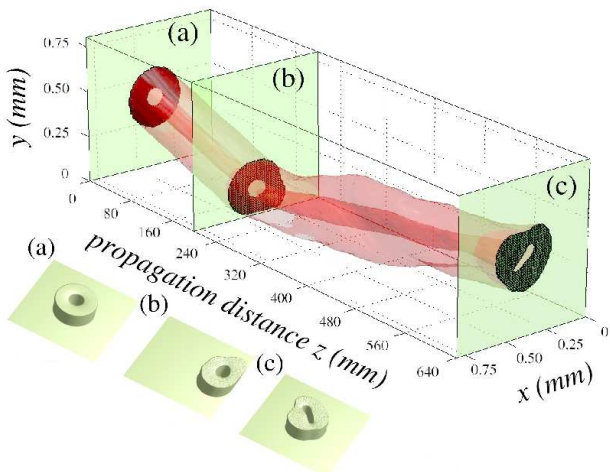}}
\caption{\label{fig6}
(Color on line) Numerical simulation of the total reflection at a planar 
boundary between a cubic-quintic and a linear media (air). For this 
stationary state the angular momentum is $l=4$
and the propagation constant $\beta=0.95\beta_{cr}$. The top image is an 
isosurface of the propagating beam. The 
boundary between the nonlinear material and
air is the plane $y=0$. The images (a)-(c) correspond to different transversal 
cutoffs (intensity profiles) of the beam at different values 
of the propagation distance $z$.}
\end{figure}

When $\beta$ is close to $\beta_{cr}$ the azimuthal 
analysis turns itself very delicate and it has to be carried out in 
a very careful way. In fact, convergence takes a much longer distance and an 
erroneous final result is obtained if the number of samples and the 
propagation step are not chosen appropriately. In this sense, combining 
the analysis with direct calculations of the splitting distance of the 
unstable eigenstates is definitively useful. In Fig. \ref{fig5}(d) it 
is zoomed the region of \ref{fig5}(c) where the perturbation for $l=3$ drops 
to zero. The points obtained propagating the eigenstates and taking the 
inverse of the distance where they split are also plotted. 
These values were subsequently scaled by the same constant value to compare 
with the perturbation eigenvalue curve. As can be appreciated, the values 
obtained from this propagation experiments fall to zero with the same 
slope as the perturbation eigenvalues do. When the stability analysis is 
not performed with enough accuracy a more steady behavior of the curve 
appears, what implies that the eigenvalue falls to zero at a higher value 
of $\beta$. This allows us to assess the validity of the perturbation 
analysis.

As a final test of the stability of the eigenstates, we have simulated 
the total reflection at a planar boundary between a cubic-quintic material 
and air for beams with different angular momenta. For the simulation we have 
used a split-step Fourier method with a $520\times 520$ grid. The idea is 
similar to the test of surface tension properties of ''liquid light 
beams'' from ref. \cite{michinel02}. As can be seen in Fig. \ref{fig6}, 
a beam with $l=4$ does not split after the total reflection, although a 
strong oscillation is observed. This is another proof of the stability of 
these nonlinear waves. We must notice that depending on the incidence angle,
a strong deformation of the beam can be induced, which can yield to a 
split or a decay of the inner vortex into several defects with lower 
topological charges \cite{paz-alonso04}.


\section{Conclusions} 

The main conclusions that can be derived from the present work are the 
following: first, stable azimuthal finite-size beams with arbitrary 
very large angular momentum can exist in optical materials with 
self-focusing (cubic) and defocussing (quintic) nonlinearity. 
Second, the shapes of these beams tend asymptotically to square-like ring 
profiles with bigger dark holes for higher values of the angular momentum. 
And finally, the critical values of the propagation constant and amplitude 
do not depend on the angular momentum of the beam.

\end{document}